\documentclass[useAMS,usenatbib,usegraphicx]{mn2e}

\def\kms{km s$^{-1}$}
\def\msun{M$_{\sun}$}
\def\rsun{R$_{\sun}$}
\def\aap{A\&A}
\def\apjl{ApJ}
\def\apj{ApJ}
\def\apjs{ApJS}
\def\aj{AJ}
\def\mnras{MNRAS}

\title[A new 40 min Period Binary WD]
{SDSS J163030.58+423305.8: A 40 minute Orbital Period Detached White Dwarf Binary\thanks{Based
on observations obtained at the MMT Observatory, a joint facility of
the Smithsonian Institution and the University of Arizona.}}

\author[M. Kilic et al.]
       {Mukremin Kilic$^{1}\thanks{Email: kilic@ou.edu}$,
       Warren R. Brown$^2$,
       J. J. Hermes$^3$,
       Carlos Allende Prieto$^{4,5}$,
       \newauthor
       S. J. Kenyon$^2$,
       D. E. Winget$^3$,
       and K. I. Winget$^3$\\
       $^1$Homer L. Dodge Department of Physics and Astronomy, University of Oklahoma,
       440 W. Brooks St., Norman, OK, 73019, USA\\
       $^2$Smithsonian Astrophysical Observatory, 60 Garden St, Cambridge, MA 02138, USA\\
       $^3$Department of Astronomy, University of Texas at Austin, RLM 16.236, Austin, TX 78712, USA\\
       $^4$Instituto de Astrof\'{\i}sica de Canarias, E-38205 La Laguna, Tenerife, Spain\\
       $^5$Departamento de Astrof\'{\i}sica, Universidad de La Laguna, E-38206 La Laguna, Tenerife, Spain
}

\begin{document}

\maketitle

\begin{abstract}

We report the discovery of a new detached, double white dwarf (WD) system with an orbital period
of 39.8 min. We targeted SDSS J163030.58+423305.8 (hereafter J1630) as part of our radial
velocity program to search for companions around low-mass WDs using the 6.5m MMT.
We detect peak-to-peak radial velocity variations of 576 \kms. The mass function and optical
photometry rule out main-sequence companions. In addition, no milli-second pulsar companions
are detected in radio observations. Thus the invisible companion is most likely another
white dwarf. Unlike the other 39 min binary SDSS J010657.39$-$100003.3, follow-up high speed
photometric observations of J1630 obtained at the McDonald 2.1m telescope
do not show significant ellipsoidal variations, indicating a higher primary mass and smaller radius.
The absence of eclipses constrain the inclination angle to $i\leq82^{\circ}$.
J1630 contains a pair of WDs, 0.3\msun\ primary + $\geq$0.3\msun\ invisible secondary,
at a separation of $\geq$0.32\rsun. The two WDs will merge in less than 31 Myr. Depending on the
core composition of the companion, the merger will form either a single core-He burning subdwarf star
or a rapidly rotating massive WD. The gravitational wave strain from J1630 is detectable by instruments like
the Laser Interferometer Space Antenna (LISA) within the first year of operation.

\end{abstract}

\begin{keywords}
        binaries: close ---
        white dwarfs ---
        stars: individual (SDSS J163030.58+423305.8) ---
        supernovae: general ---
        gravitational waves
\end{keywords}

\section{INTRODUCTION}

Radial velocity surveys of WDs are the best way to identify short period systems that may merge within
a Hubble time and produce normal Type Ia or underluminous ``.Ia'' supernovae
\citep[SNe,][]{webbink84,iben84,bildsten07}. However, even large spectroscopic surveys
like the SN Ia Progenitor Survey (SPY) have limited success in identifying merger systems
\citep{napiwotzki01,nelemans05}.
Many of the radial velocity variables in SPY and other surveys turn out to be low-mass WDs
\citep{marsh95,maxted00,napiwotzki07}. Only 24 double WD systems were known in 2005,
and all but three have primary WDs with $M\leq$0.5\msun \citep{nelemans05}.

The Galaxy is not old enough to produce low-mass WDs through single star evolution, but they can form
in unusual systems \citep[see][for a review]{kilic07b,nelemans98}. The majority of low-mass WDs are found
in binary systems \citep{brown11a}, where the progenitor main-sequence stars
experience enhanced mass-loss during one or two common-envelope phases.
Hence, low-mass WDs are prime targets for discovering short period merger systems.

\citet{kilic10,kilic11a} and \citet{brown10} have established a radial velocity program,
the ELM Survey, to search for companions around known extremely low-mass ($M\sim$0.2\msun)
WDs in the SDSS Data Release 7 footprint and
the MMT Hypervelocity Star Survey \citep{brown06}. So far, the ELM Survey has identified 12 WD merger systems,
tripling the number of systems known \citep{kilic11a}. Depending on the unknown mass ratios,
these systems may be the progenitors of stable mass-transfer AM CVn binaries, single helium-rich subdwarfs,
extreme helium stars, or underluminous SNe. The ELM Survey is also responsible for finding the first tidally distorted
WD J0106 \citep{kilic11b} and the 12 min orbital period detached, eclipsing binary WD system J0651 \citep{brown11c}.
These two systems, J0106 and J0651, are currently the only known detached binary WD systems with periods shorter
than 1 hr. All other known binary WD systems with $P<$ 1 hr are in interacting AM CVn systems.

Here we present the discovery of a new detached binary WD system with $P<$ 1 hr found in the ELM
Survey. J1630 was recognized as a low-mass WD in the SDSS DR4 WD catalog \citep{eisenstein06}.
Our follow-up radial velocity observations demonstrate that J1630 contains a pair of WDs with
an orbital period of only 39.8 minutes.
In Section 2 we describe our spectroscopic and photometric observations. 
In Sections 3 and 4 we constrain the physical parameters of the binary and discuss
the nature and future evolution of J1630. We summarize our conclusions in Section 5.

\section{OBSERVATIONS}

We used the 6.5m MMT with the Blue Channel spectrograph to obtain medium resolution spectroscopy
of J1630 on UT 2011 March 1 and April 22-23.
We operate the spectrograph with the 832 line mm$^{-1}$
grating in second order, providing wavelength coverage from 3600 \AA\ to 4500 \AA\ and a spectral
resolution of 1.2 \AA. J1630 has $g=19.0$ mag. We started our observations with 400 s exposures,
but reduced the exposure time to 300 s after detecting significant velocity variations in back-to-back
exposures. We obtain all observations at the parallactic angle, with a comparison
lamp exposure paired with every observation. We flux-calibrate using blue
spectrophotometric standards \citep{massey88}, and we measure radial velocities
using the entire spectrum in the range 3700-4430 \AA\ and the cross-correlation package RVSAO.
The details of our data reduction procedures are discussed in \citet{kilic10}.

Realizing that J1630 has a period similar to the tidally distorted WD J0106, we
acquired high speed photometric observations of J1630 to search for eclipses and ellipsoidal
variations. We used the McDonald 2.1m Otto Struve Telescope with the Argos frame 
transfer camera \citep{mukadam05} over 1 night in 2011 May and 3 nights in 2011 July.
We obtained 1819 images of J1630 with a 1mm BG40 filter and 15 or 30 s exposures for a total exposure
time of 9.6 hr.

\section{RESULTS}

\subsection{The Orbital Period}

Table 1 lists our radial velocity measurements for J1630.
We compute best-fit orbital elements using the code of \citet{kenyon86}.
To verify the uncertainty estimates, we perform a Monte Carlo analysis using
10000 sets of simulated radial velocities \citep[see][]{brown10}. We adopt the inter-quartile range in the
period and orbital elements as the uncertainty.

\begin{table}
\centering
\caption{Radial Velocity Measurements for J1630}
\begin{tabular}{cr}
\hline 
HJD$-$2455600 & $v_{helio}$ \\
(days) & (\kms) \\
\hline
22.034532 &    201 $\pm$ 22 \\
74.838943 &   $-$1 $\pm$ 22 \\
74.843781 & $-$300 $\pm$ 33 \\
74.849614 & $-$274 $\pm$ 24 \\
74.854429 &     60 $\pm$ 53 \\
74.859313 &    310 $\pm$ 46 \\
74.965241 &    123 $\pm$ 16 \\
74.968898 &    246 $\pm$ 16 \\
74.972579 &    194 $\pm$ 25 \\
74.976236 &  $-$20 $\pm$ 26 \\
74.981699 & $-$262 $\pm$ 15 \\
74.985380 & $-$280 $\pm$ 18 \\
74.989038 & $-$138 $\pm$ 14 \\
74.992695 &     47 $\pm$ 39 \\
74.996514 &    278 $\pm$ 39 \\
74.998806 &    314 $\pm$ 39 \\
75.912767 &    255 $\pm$ 20 \\
75.916447 &    160 $\pm$ 29 \\
75.920093 & $-$129 $\pm$ 23 \\
75.924700 & $-$298 $\pm$ 24 \\
75.928380 & $-$125 $\pm$ 31 \\
\hline
\end{tabular}
\end{table}

\begin{figure}
\includegraphics[width=2.65in,angle=-90]{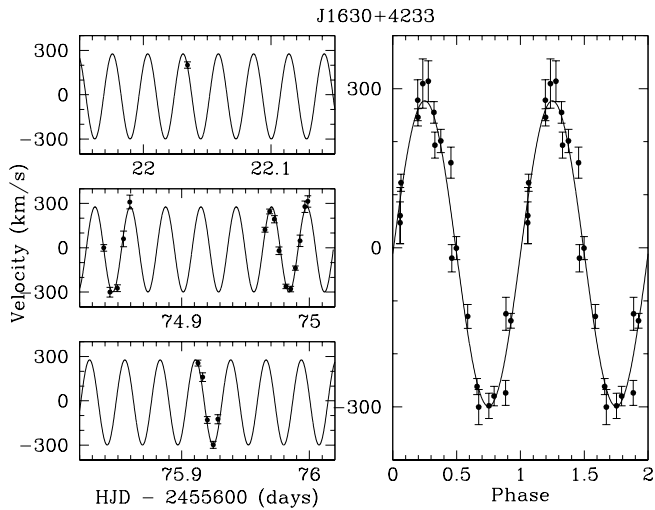}
\caption{The heliocentric radial velocities of J1630 observed over three nights in 2011 March and April
(left panels). The right panel shows all of these data points phased with the best-fit
period. The solid line represents the best-fit model for a circular orbit with a period
of 39.8 min and $K=288.1$ \kms.}
\end{figure}

J1630 exhibits radial velocity variations with a semi-amplitude of $K = 288.1 \pm 4.9$
\kms and orbital period of $P = 0.027659 \pm 0.000043$ d, or 39.8 $\pm$ 0.1 min.
Figure 1 shows the best-fit orbit compared to the observed radial velocities.
Due to our relatively long exposure times, our data do not sample the
velocity curve well near maximum and minimum. Thus, we underestimate the
velocity semi-amplitude. To correct for this underestimate, we sample a sine
curve at the exact 21 phases of our observations with $P/8$ long integrations. We recover the exact
period, but $K$ is systematically underestimated by 2.7\%. Thus, the corrected velocity semi-amplitude
is $K= 295.9$ \kms. With this correction,
J1630 has a mass function of $f= 0.07423 \pm 0.00369$ \msun.
The systemic velocity (after subtracting the gravitational redshift of 7.6 \kms) is $-$18.7 $\pm$ 3.8
\kms\ and the time of spectroscopic conjunction is HJD 2455622.024130 $\pm$ 0.000097 d.

\subsection{The Physical Parameters of the Binary}

\citet{eisenstein06} derive $T_{\rm eff} = 14850 \pm 360$ K and $\log{g} = 6.89 \pm 0.13$
from the low-quality SDSS spectrum of J1630. \citet{kilic07a} find $T_{\rm eff} = 14180 \pm
940$ K and $\log{g} = 7.08 \pm 0.07$ based on a re-analysis of the same spectrum.
Our higher resolution and higher signal-to-noise MMT spectra
provide a better measurement of the atmospheric parameters of the visible WD.
We perform stellar atmosphere model fits using synthetic DA WD spectra kindly provided by D.\ Koester.

Figure 2 shows the composite spectrum and our fits using the entire spectrum (top panel)
and also using only the Balmer line profiles (bottom left panel). To derive robust statistical errors,
we also perform fits to the individual spectra.
We find a best-fit solution of $T_{\rm eff} = 14670 \pm 320$ K and $\log{g} = 7.05 \pm 0.08$ from the composite
spectrum. Fitting only the Balmer lines, we obtain $T_{\rm eff} = 15560 \pm 590$ K and $\log{g} = 6.93 \pm 0.05$.
These results are consistent with each other and the previous $T_{\rm eff}$ and $\log{g}$ estimates within the errors.
The best-fit model using the continuum shape provides a slightly better fit to the SDSS photometry (Figure 2) and we adopt that
model for the remainder of the paper. We note that all of the studies described above are done using
\citet{koester10} models. \citet{gianninas11} find a systematic offset of $\log{g} \approx 0.1$ dex between their analysis
and the analysis done using \citet{koester10} models. Hence, the systematic errors in our $\log{g}$ measurement are $\approx0.1$ dex.

\begin{figure}
\includegraphics[width=3.4in]{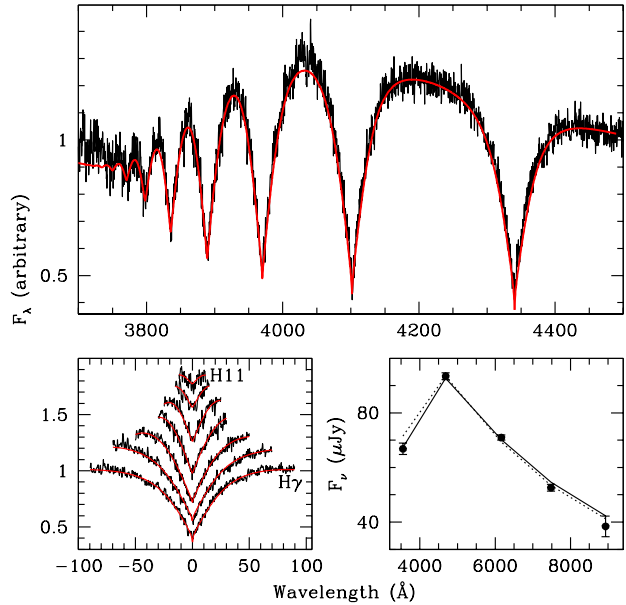}
\caption{Model fits (red lines) to the composite spectrum (jagged line, top panel)
and to the Balmer line profiles of J1630 (bottom left panel).
The spectral energy distribution of J1630 (filled circles, bottom right panel)
compared to the best-fit model using the continuum shape (solid line) and
using only the Balmer line profiles (dotted line).}
\end{figure}

Based on the improved \citet{panei07} tracks \citep[see][]{kilic10} for ELM WDs,
the visible WD in J1630 is a 160 Myr old 0.30 $\pm$ 0.02 \msun\ WD with a
radius $R= 0.025 \pm 0.003$ \rsun. Its absolute
magnitude $M_g = 9.8$ corresponds to a distance of 0.7 kpc. Based on five epochs from
the USNO-B and the SDSS, \citet{munn04} measure a proper motion of 
($\mu_{\alpha} cos \delta, \mu_{\delta}) = (2.3 \pm 3.5, -7.3 \pm 3.5$) mas yr$^{-1}$.
J1630 is 0.5 kpc below the Galactic plane and it has $U= 29 \pm 11 , V= -13 \pm 9$,
and $W= -11 \pm 9$ km s$^{-1}$ with respect to the local standard of rest \citep{hogg05}.
J1630 is a disk star.

The mass function implies a companion mass of $\geq$ 0.30 \msun.
Such a main-sequence companion would have been detected in the SDSS photometry.
Therefore, the companion is a compact object. Based on the mass function alone,
the probability of a neutron star (1.4-3.0 \msun) companion is 5\%. However,
no milli-second pulsar companions are detected in the radio \citep{agueros09}.
Hence, the companion is very likely another WD.

\subsection{The Light Curve}

Figure 3 shows the fourier transform (FT) of the J1630 light curve. The light curve
is affected by low frequency sky noise due to the changes in the atmosphere
and the color differences between the WD and the reference stars. The middle panel
shows the FT pre-whitened by the main peak in the top panel. There are many aliases, but
the highest peak is at the orbital period (dashed line). The FT for the brightest comparison star
(bottom panel) does not show any significant variations at the same frequency.

\begin{figure}
\includegraphics[width=3.4in]{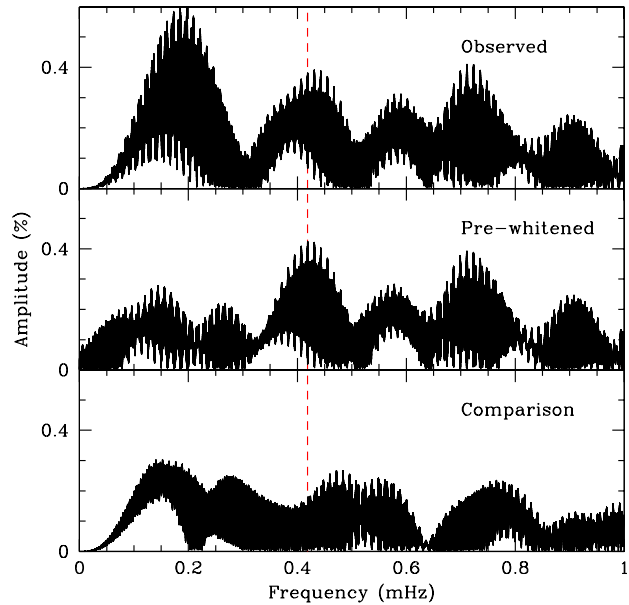}
\caption{Fourier transform of the J1630 light curve (top panel). The data shows low-frequency
sky noise at $\sim$0.2 mHz. After pre-whitening, the strongest peak is at the orbital period
of the binary (dashed line). No significant peak is observed at the same frequency for the comparison star.}
\end{figure}

Figure 4 shows the Argos light curve of J1630 (top panel) folded over the best-fit
orbital period of 39.8 min. The bottom panel shows the same light curve binned into
100 orbital phase bins. Unlike J0106 and J0651, J1630 does not show any significant ellipsoidal
variations. The amplitude of the ellipsoidal effect is roughly
$\delta f_{ell} = (m_2/m_1)(r_1/a)^3$, where $a$ is the orbital semi-major axis and $r_1$
is the radius of the primary \citep{zucker07,shporer10}.
Even though J0106 and J1630 have almost the same orbital period,
J1630 is about a factor of two more massive and a factor of two smaller in size compared to J0106.
Therefore, the expected amplitude of the ellipsoidal variations is $<$0.1\%, consistent with
the observations.

\begin{figure}
\includegraphics[width=3.4in]{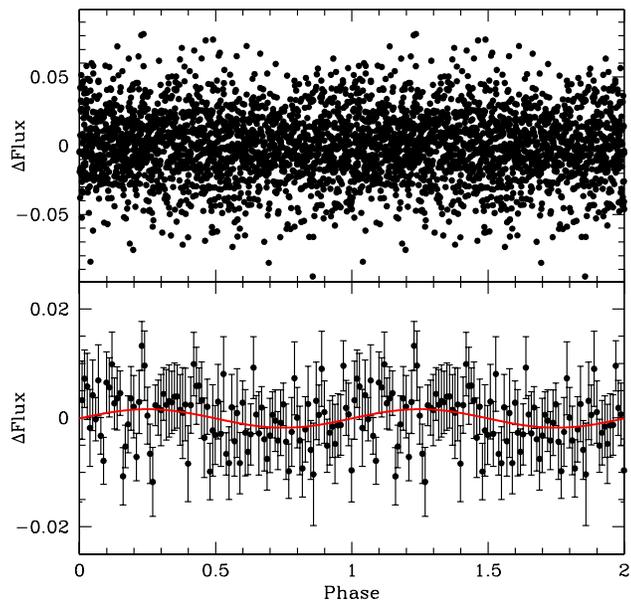}
\caption{High speed photometry of J1630 over 9.6 hours (top
panel). The bottom panel shows the same light curve binned into 100 phase bins.
The solid line shows a 0.17\% amplitude signal at the orbital period.}
\end{figure}

We expect to see $\approx$0.24\% amplitude sinusoidal variations due to the relativistic beaming effect.
Fitting a sinusoid, the lightcurve is best explained by 0.17\% $\pm$ 0.05\% variations at the orbital
period (see the bottom panel in Figure 3). These variations are consistent with the expected beaming signal.
To verify that the observed variations in the light curve are not a statistical fluctuation,
we perform a bootstrap analysis. We create 10000 synthetic light curves by random sampling with replacement
from the observed light curve. This analysis shows that the probability of measuring
a random signal with $\geq$0.17\% amplitude at the orbital period is 2\%. 
Based on Figures 3 and 4 and our statistical analysis, the doppler boosting signal is likely detected
in J1630. However, the boosting signal does not provide any new physical constraints on the properties of
this binary.

\section{DISCUSSION}

Along with J0106 and J0651, J1630 is only the third detached WD binary known to have a period shorter than an hour.
All three were discovered in the last year as part of the ELM Survey. Previously, all known systems with $P<1$ hr
were interacting AM CVn systems. We do not see any evidence of mass transfer (no emission lines and no obvious excess
continuum) in the three targets mentioned above.
The primary WDs in J0106 and J0651 both show ellipsoidal variations due to their larger size and shorter orbital periods.
These variations are extremely useful for constraining the inclination angle of the systems. However, J1630 does not
show any significant ellipsoidal variations, and we only have an upper limit on the inclination angle due to the lack
of eclipses. 

For the companion to avoid detection in the SDSS photometry and our spectroscopy implies that it is
$\ge10\times$ fainter than the visible WD.
For an edge on orbit, such a companion would have $M=0.30$ \msun,
$T_{\rm eff}\leq7500$ K, and $R\leq0.021$ \rsun\ \citep{panei07}. A total eclipse would be $\leq$70\% deep and last
for about 1.8 minutes.
The lack of eclipses in the photometry constrain the inclination angle to $i\leq82^{\circ}$.
Hence, J1630 is best explained by a binary system containing a 0.30 \msun\ WD with a $M\geq$0.30 \msun\ WD
companion at a separation of $\geq$0.32 \rsun.

The two WDs in the J1630 binary will merge in $\leq$31 Myr due to gravitational wave radiation \citep{lan58}.
After J0651, J1630 is currently the second quickest WD merger system known.
When the mass transfer starts, J1630 will most likely have unstable mass transfer due to the mass ratio of its
components being close to unity \citep{marsh04}. However, the merger outcome is uncertain because of the unknown
inclination angle and the companion mass. If the companion is another He-core WD, the merger will likely
create a single He-burning subdwarf in 23-31 Myr. If the companion is a more massive carbon/oxygen core WD,
the system will merge in $\leq$23 Myr to form a rapidly rotating massive WD \citep[see][for other possibilites]{kilic10}.

Short period binary WDs are important gravitational wave sources. For example, the 12 minute orbital period binary
J0651 should be detected by LISA within its first week of operation
\citep{brown11c,nelemans04}. Similarly, the previously discovered 39 min orbital period system J0106 may be detected
by LISA after 1 yr of observations. With an orbital period similar to J0106,
J1630 is also a promising candidate for detection. For an average inclination angle of 60$^{\circ}$ and model-dependent
distance of 0.7 kpc, we expect the gravitational wave strain at Earth $\log h = -22.0$ at a frequency
$\log \nu$ (Hz) = $-3.08$ \citep{roelofs07}. This is at the S/N = 5 detection limit of LISA after 1 year of observations.

\section{CONCLUSIONS}

We report the discovery of a new 40 min orbital period detached binary WD system, J1630.
Along with the previously discovered 12 min and 39 min orbital period systems, J0651 and J0106,
J1630 is only the third known detached binary WD system with a period less than an hour.
All three systems are excellent gravitational wave sources to be detected by LISA.

J1630 is almost a twin of the 39 min orbital period system J0106. However, due to its smaller size
and larger mass, significant ellipsoidal variations are neither expected nor detected in its light
curve. Therefore, the inclination angle of the system cannot be constrained accurately with the
current data. Follow-up high-speed photometric observations at a larger telescope will be useful
to constrain the doppler boosting signal more accurately and to search for grazing eclipses.
The two WDs in this system will merge in less than 31 Myr and likely form a single subdwarf star
or a rapidly rotating massive WD.

\section*{Acknowledgements}

We thank D. Koester for kindly providing WD model spectra.
DEW and JJH gratefully acknowledge the support of the NSF
under grant AST-0909107 and the Norman Hackerman Advanced
Research Program under grants 003658-0255-2007 and 
003658-0252-2009.

\end{document}